\documentclass[aps,prc,preprint,superscriptaddress,showpacs,floatfix]{revtex4}
\usepackage{graphicx,color}
\usepackage{braket}
\usepackage{amsmath}
\usepackage{amssymb}
\newlength{\figwidth}

\begin{document}
\setlength{\figwidth}{0.98\columnwidth}

\title{Effects of pairing correlation on low-lying quasi-particle resonance in neutron drip-line nuclei}

\author{Yoshihiko Kobayashi}
\affiliation{Graduate School of Science and Technology, Niigata University, Niigata 950-2181, Japan}
\author{Masayuki Matsuo}
\affiliation{Department of Physics, Faculty of  Science, Niigata University, Niigata 950-2181, Japan}

\begin{abstract}
We discuss effects of pairing correlation on quasi-particle resonance. We analyze in detail how the width of low-lying ($E_{x}\lesssim 1~\mathrm{MeV}$) quasi-particle resonance is governed by the pairing correlation in the neutron drip-line nuclei. We consider the ${}^{46}$Si + n system to discuss low-lying $p$ wave quasi-particle resonance. Solving the Hartree-Fock-Bogoliubov equation in the coordinate space with scattering boundary condition, we calculate the phase shift, the elastic cross section, the resonance width and the resonance energy. We found that the pairing correlation has an effect to {\it reduce} the width of quasi-particle resonance which originates from a particle-like orbit in weakly bound nuclei.
\end{abstract}

\maketitle

\section{Introduction}
Weakly bound nuclei near the drip-line have properties which are not seen in strongly bound stable nuclei. The neutron halo is an typical example~\cite{Tanihata1985,Tanihata2013}. Apart from quantal penetration caused by the small separation energy, the neutron pairing correlation plays crucial roles here, for example, to determine the binding of two-neutron halo nuclei~\cite{Hansen1987,Bertsch1991,Meng1996,Barranco2001,Myo2002}. Note, however, that the pairing correlation in weakly bound nuclei is different from that in stable nuclei since it causes configuration mixing involving both bound and unbound (continuum) single-particle orbits, and this continuum coupling brings about novel features~\cite{Meng1996,Meng2006,Dobaczewski1996,Dobaczewski2007,Bennaceur2000,Hamamoto2003,Hamamoto2004,Matsuo2005,Matsuo2010,Zhang2011,Zhang2014}. For example, the pairing correlation persists in drip-line nuclei only with the continuum coupling  to allow binding of a two-neutron halo~\cite{Meng1996,Meng2006}. The continuum coupling is necessary also for the di-neutron correlation, characteristic spatial correlation in neutron-rich nuclei~\cite{Matsuo2005,Hagino2005,Pillet2007,Zhang2014}. On the other hand, the continuum coupling has seemingly opposite mechanism to suppress the development of the halo radius, called the pairing anti-halo effects~\cite{Bennaceur2000,Hamamoto2004,Chen2014}.

Another interesting example is possible manifestation of a new type of resonance generated by the pairing correlation and the continuum coupling, called the quasi-particle resonance~\cite{Belyaev1987,Bulgac1980}. If one describes a single-particle scattering problem within the scheme of Bogoliubov's quasi-particle theory, even a scattering state becomes a quasi-particle state which has both `particle' and `hole' components. In other words, an unbound nucleon couples to a Cooper pair and a bound hole orbit, then forms a resonance. This quasi-particle resonance is expected also to exhibit new features in weakly bound nuclei since the continuum coupling becomes stronger as the separation energy decreases.

In the case of well bound stable nuclei, the depth of Fermi surface is around 8 MeV. Therefore, quasi-particle resonances, which emerge above the separation energy, have excitation energy larger than 8 MeV, and hence they correspond to deep hole orbits. The excitation energy $E^{\mathrm{stable}}_{x}$ of quasi-particle resonance is much larger than the pair gap $\Delta$: $E^{\mathrm{stable}}_{x}\gg\Delta$. In this case, the effect of the pairing correlation is treated in a perturbative way~\cite{Belyaev1987,Bulgac1980}. The resonance width $\Gamma$, for example, is evaluated on the basis of the Fermi's golden rule. The width $\Gamma$ is predicted to be proportional to the square of the pair gap $|\Delta_{\mathrm{average}}|^{2}$, and $\Gamma$ is estimated to be small (i.e. order of 1-100 keV)~\cite{Belyaev1987}, much smaller than the experimentally known typical width (several MeV) of the deep-hole resonances~\cite{RingSchuck}. Experimental identification of the pairing effect on the deep-hole resonances is not very promising in this respect~\cite{Dobaczewski1996}.

In the case of small separation energy, in particular, in neutron-rich nuclei, property of the quasi-particle resonance may be different from those in stable nuclei. A neutron-rich nucleus has a shallow Fermi energy, with an extreme depth smaller than 1 MeV to be realized in neutron drip-line nuclei. In this case the excitation energy of a quasi-particle resonance might be comparable with or smaller than the pair gap: $E^{\mathrm{unstable}}_{x}\lesssim\Delta$. The pairing correlation may cause strong configuration mixing between weakly bound orbits and low-lying continuum orbits, since both are located near the Fermi surface. The perturbative description may not be applicable, and we expect undisclosed relation between the quasi-particle resonance and the pairing correlation.

The small neutron separation energy provides another merit in studying the quasi-particle resonance. In this case the quasi-particle resonance appears also in the low-lying region where the level density is low. Other mechanisms beyond the mean-field approximation, for instance, the fragmentation due to coupling to complex configurations~\cite{Bertsch1983}, are expected to be suppressed. This might increase the possibility to observe the quasi-particle resonance directly.

There exist several theoretical works that studied quasi-particle resonance in nuclei near the neutron drip-line~\cite{Hamamoto2003,Hamamoto2004,Zhang2011,Grasso2000,Fayans2000,Michel2008,Pei2011,Zhang2012,Oba2009,Zhang2013,Sandulescu2000,Betan2006,Sandulescu2005}. Many of them employ the selfconsistent Hartree-Fock-Bogoliubov (HFB) scheme~\cite{Zhang2011,Grasso2000,Fayans2000,Michel2008,Pei2011,Zhang2012}, or its variation in which the Hartree-Fock potential is replaced with the Woods-Saxon potential~\cite{Hamamoto2003,Hamamoto2004}. The quasi-particle resonance in deformed nuclei is also discussed~\cite{Oba2009,Zhang2013}. Approximate schemes using the Hartree-Fock+BCS theory are also adopted both in non-relativistic and relativistic frameworks~\cite{Sandulescu2000,Betan2006,Sandulescu2005}. Despite these previous studies, effects of the pairing correlation on the low-lying quasi-particle resonance in weakly bound nuclei have not been revealed yet. We shall discuss this subject in order to understand behavior of the pairing correlation in drip-line nuclei and unbound nuclei.

In the present study, we particularly aim to reveal effects of the pairing correlation on the width of low-lying quasi-particle resonance in drip-line nuclei. We focus on neutron resonances, in particular, in the $p$ wave having small excitation energy $E_{x}\lesssim$ a few MeV. The continuum coupling is expected to be influential for neutrons in low angular momentum partial waves, i.e. in the $s$ and $p$ waves because of no (or small) Coulomb and centrifugal barriers. And neutrons in these partial waves plays an important role in the neutron halo. Also a scattering neutron in low angular momentum waves is a major contributor in the low-energy neutron capture phenomena~\cite{Raman1985}, important for the astrophysical applications. In the present work, we discuss the $p$ wave quasi-particle resonance as a first step of a series study. The case of $s$ wave, which involves a virtual state, will be discussed separately in a future publication.

It is not appropriate to treat effects of the pairing correlation as a perturbation in the calculation of the resonance width in weakly bound nuclei. We therefore describe the continuum quasi-particle states by solving numerically the Hartree-Fock-Bogoliubov equation (equivalent to the Bogoliubov de-Genne equation) in the coordinate space~\cite{Dobaczewski1996,Dobaczewski1984,Belyaev1987,Bulgac1980} to obtain the wave function of a neutron quasi-particle in the continuum. We impose the scattering boundary condition~\cite{Belyaev1987,Hamamoto2003,Hamamoto2004,Grasso2000}. In this way, we calculate the phase shift for the continuum quasi-particle state and the elastic cross section for a neutron scattered by the superfluid nucleus. Then the resonance width and the resonance energy are extracted from the obtained phase shift. As a concrete example, we describe ${}^{46}$Si and an impinging neutron, in other words, a quasi-particle resonance in ${}^{47}$Si. Hartree-Fock-Bogoliubov calculations predict that this nucleus is located at or close to the neutron drip-line~\cite{Stoitsov2003}. Also it has the neutron $2p$ orbits in ${}^{46}$Si are expected to be weakly bound or located just above the threshold energy.

This paper is constructed as follows: In Sect.~2, we explain the HFB theory in the coordinate space, the scattering boundary condition of the Bogoliubov quasi-particle and some details of the adopted model. In Sect.~3, we show the results of numerical analysis performed for the ${}^{46}$Si + n system. We also discuss effects of the pairing correlation on the resonance width using systematic calculation with various pairing strengths and nuclear potential depths. Finally, we draw conclusions in Sect.~4.

\section{Theoretical Framework}
\subsection{The Hartree-Fock-Bogoliubov equation in the coordinate space with the scattering boundary condition}
We introduce the wave function of the Bogoliubov quasi-particle state in the notation of Ref.~\cite{Dobaczewski1984,Matsuo2001}. It has two components;
\begin{equation}
\phi_{i}(\vec{r}\sigma)=
\left(
\begin{array}{c}
\varphi_{1,i}(\vec{r}\sigma) \\
\varphi_{2,i}(\vec{r}\sigma)
\end{array}
\right).
\end{equation}
Here $\vec{r}$ is the spatial coordinate and $\sigma$ represents the spin variable. Assuming that the system has spherical symmetry, we write the Bogoluibov quasi-particle wave function as
\begin{equation}
\varphi_{1,i}(\vec{r}\sigma)=\frac{u_{lj}(r)}{r}[Y_{l}(\theta ,\varphi)\chi_{\frac{1}{2}}(\sigma)]_{jm},\quad \varphi_{2,i}(\vec{r}\sigma)=\frac{v_{lj}(r)}{r}[Y_{l}(\theta ,\varphi)\chi_{\frac{1}{2}}(\sigma)]_{jm},
\end{equation}
where $l$, $j$ and $m$ are the angular momentum quantum numbers of the quasi-particle state, with $Y$ and $\chi$ being the spherical harmonics and the spin wave function. We also assume that the HF potential and the pair hamiltonian $\Delta (\vec{r})$ are local and real, then the Hartree-Fock-Bogoliubov equation in the coordinate space is written as
\begin{equation}
\left(
\begin{array}{ccc}
-\frac{\hbar^{2}}{2m}\frac{d^{2}}{dr^{2}}+U_{lj}(r)-\lambda & \Delta (r) \\
\Delta (r) &\frac{\hbar^{2}}{2m}\frac{d^{2}}{dr^{2}}-U_{lj}(r)+\lambda
\end{array}
\right)
\left(
\begin{array}{c}
u_{lj}(r) \\
v_{lj}(r)
\end{array}
\right)
=E
\left(
\begin{array}{c}
u_{lj}(r) \\
v_{lj}(r)
\end{array}
\right),
\label{hfbeq2_sec4}
\end{equation}
where $\lambda (<0)$ and $E$ are the Fermi energy and the quasi-particle energy, respectively. Here the upper component of quasi-particle wave function $u_{lj}(r)$ represents an amplitude of the quasi-particle having the particle character, called hereafter the `particle' component in short. The lower component $v_{lj}(r)$ represents the `hole' component. $U_{lj}(r)$ is the mean field potential and $m$ is the mass of neutron. The spectrum of quasi-particle consists of discrete states with $E<|\lambda|$ and continuum states with $E>|\lambda|$~\cite{Dobaczewski1984}.

We intend to describe a system consisting of a superfluid nucleus and an impinging neutron, which in principle should be treated as a many-body unbound state. However, we adopt an approximation to which the neutron is treated as an unbound quasi-particle state, governed by Eq.~(3), built on the pair correlated even-even nucleus. In other words, we neglect selfconsistent effect of unbound neutron on the mean field and the pair correlation. Under this assumption, we focus on continuum quasi-particle states with $E>|\lambda|$ which correspond to unbound single-particle states with positive neutron kinetic energy. We impose the scattering boundary condition on the Bogoliubov quasi-particle at distances far outside the nucleus as
\begin{equation}
\frac{1}{r}\left(
\begin{array}{c}
u_{lj}(r)\\
v_{lj}(r)
\end{array}
\right)=
C\left(
\begin{array}{c}
\cos \delta_{lj}j_{l}(k_{1}r)-\sin \delta_{lj}n_{l}(k_{1}r)\\
Dh^{(1)}_{l}(i\kappa_{2}r)
\end{array}
\right)\xrightarrow[r\to\infty]{}
C\left(
\begin{array}{c}
\frac{\sin\left( k_{1}r-\frac{l\pi}{2}+\delta_{lj} \right)}{k_{1}r}\\
0
\end{array}
\right)
\label{bogoscatt}
\end{equation}
where $k_{1}=\sqrt{2m(\lambda+E)}/\hbar$, $\kappa_{2}=\sqrt{-2m(\lambda-E)}/\hbar$~\cite{Belyaev1987,Bulgac1980,Dobaczewski1984,Grasso2000,Hamamoto2003}. The normalization factor $C$ is $C=\sqrt{2mk_{1}/\hbar^{2}\pi}$ to satisfy $\sum_{\sigma}\int d\vec{r}\phi^{\dagger} (\vec{r}\sigma,E)\phi (\vec{r}\sigma,E^{\prime})=\delta(E-E^{\prime})$. Here $\delta_{lj}$, $j_{l}(z)$, $n_{l}(z)$, $h^{(1)}_{l}(z)$ are the phase shift, the spherical Bessel function, the spherical Neumann function and the first kind spherical Hankel function, respectively. The quasi-particle resonance can be seen in the elastic scattering of a neutron, and the elastic cross section $\sigma_{lj}$ associated with each partial wave is
\begin{eqnarray}
\sigma_{lj}=\frac{4\pi}{k^{2}_{1}} \left( j+\frac{1}{2} \right) \sin ^{2}\delta_{lj}.
\end{eqnarray}

\subsection{Details of numerical calculation}
We solve the radial HFB equation~(\ref{hfbeq2_sec4}) in the radial coordinate space  under the scattering boundary condition~(\ref{bogoscatt}) of the Bogoliubov quasi-particle. In the present study, we simplify the HF mean field by replacing it with the Woods-Saxon potential in a standard form:
\begin{equation}
U_{lj}(r)=\left[ V_{0}+ (\vec{l}\cdot\vec{s})V_{\mathrm{SO}}\frac{r^{2}_{0}}{r}\frac{d}{dr}  \right] f_{\mathrm{WS}}(r)+\frac{\hbar^{2}l(l+1)}{2mr^{2}}, \quad f_{\mathrm{WS}}(r)=\left[ 1+\exp \left( \frac{r-R}{a} \right) \right]^{-1}.
\label{hamil}
\end{equation}
Although the selfconsistency of the mean fields is neglected, an advantage of this treatment is that we can easily change parameters of the potentials, facilitating systematic numerical analysis. On the other hand, effects of weakly binding on the potential, for instance, large diffuseness and  long tail, are not taken into account in the present calculation. We also assume that the pair potential $\Delta (r)$ has the Woods-Saxon shape:
\begin{equation}
\Delta(r)=\Delta_{0}f_{\mathrm{WS}}(r),
\end{equation}
following Ref.~\cite{Hamamoto2003}. The magnitude of the pair potential $\Delta_{0}$ is controled by the average pair strength $\bar{\Delta}$~\cite{Hamamoto2003}:
\begin{equation}
\bar{\Delta}=\frac{\int^{\infty}_{0}r^{2}\Delta (r)f_{\mathrm{WS}}(r)dr}{\int^{\infty}_{0}r^{2}f_{\mathrm{WS}}(r)dr}=0.0 - 3.0~\mathrm{MeV}.
\end{equation}
We change the strength $\bar{\Delta}$ from 0.0 MeV to 3.0 MeV in this study, considering the empirical systematics of the pair gap $\Delta\sim 12.0/\sqrt{A}$ MeV~\cite{BohrMottelson} ($\Delta\sim 1.7$ MeV for ${}^{46}$Si). The parameters of the Woods-Saxon potential are taken from Ref.~\cite{BohrMottelson}. The radial wave function is numerically solved up to $r_{\mathrm{max}}=40$ fm, where it is connected to the Hankel functions, Eq.~(4).

We consider the ${}^{46}$Si + n system for the following reasons. First, ${}^{46}$Si is predicted be the drip-line nucleus in Si isotopes and the deformation of this nucleus is small according to the HFB calculations (for instance, the Refs~\cite{Stoitsov2003,Werner1996,Terasaki1997}). It may be reasonable to assume that ${}^{46}$Si has spherical shape in the present calculation. Second, the neutron $2p_{3/2}$ or $2p_{1/2}$ orbits are expected be either weakly bound or slightly unbound, and hence they are expected to form low-lying quasi-particle resonances. Note that ${}^{46}$Si has not been observed yet experimentally~\cite{Thoennessen2012}.

The neutron single-particle energies around the Fermi energy for ${}^{46}$Si in the Woods-Saxon potential is shown in  Table.~1. Both of $2p$ orbits are bound very weakly for the original parameter set. In particular, the energy of $2p_{1/2}$ orbit is very small: $e_{\mathrm{sp}}=-0.056$ MeV. For the Fermi energy $\lambda$, we use a fixed value $\lambda=-0.269$ MeV which is obtained by the Woods-Saxon-Bogoliubov calculation~\cite{Oba2009}.
\begin{table}[t]
  \centering
  \begin{tabular}{ccc} \hline
    Single-particle orbit && Single-particle energy $e_{\mathrm{sp}}$ [MeV] \\ \hline
    $2p_{1/2}$ && -0.056  \\
    $2p_{3/2}$ && -1.068  \\
    $1f_{7/2}$ && -2.821  \\ \hline
  \end{tabular}
\caption{Neutron single-particle orbits in the Woods-Saxon potential of ${}^{46}$Si, obtained with the standard Woods-Saxon parameter~\cite{BohrMottelson}.}
\label{spene}
\end{table}

\section{Results and discussion}
\subsection{Cross section and phase shift of neutron elastic scattering}
Figure~1 shows the calculated elastic cross section which is obtained (a) without the pairing correlation ($\bar{\Delta}=0.0$ MeV) and (b) with the pairing correlation ($\bar{\Delta}=1.0$ MeV). In the case of $\bar{\Delta}=0.0$ MeV, single-particle potential resonances are found in the $f_{5/2}$ and $g_{9/2}$ waves, corresponding to the $1f_{5/2}$ and $1g_{9/2}$ orbits trapped by the centrifugal barrier. Note that configurations with the last neutron occupying the $2p_{3/2}$ or $2p_{1/2}$ orbits are bound states, and are not seen in Fig.~1~(a). 
\begin{figure}
\begin{center}
\includegraphics[scale=0.59]{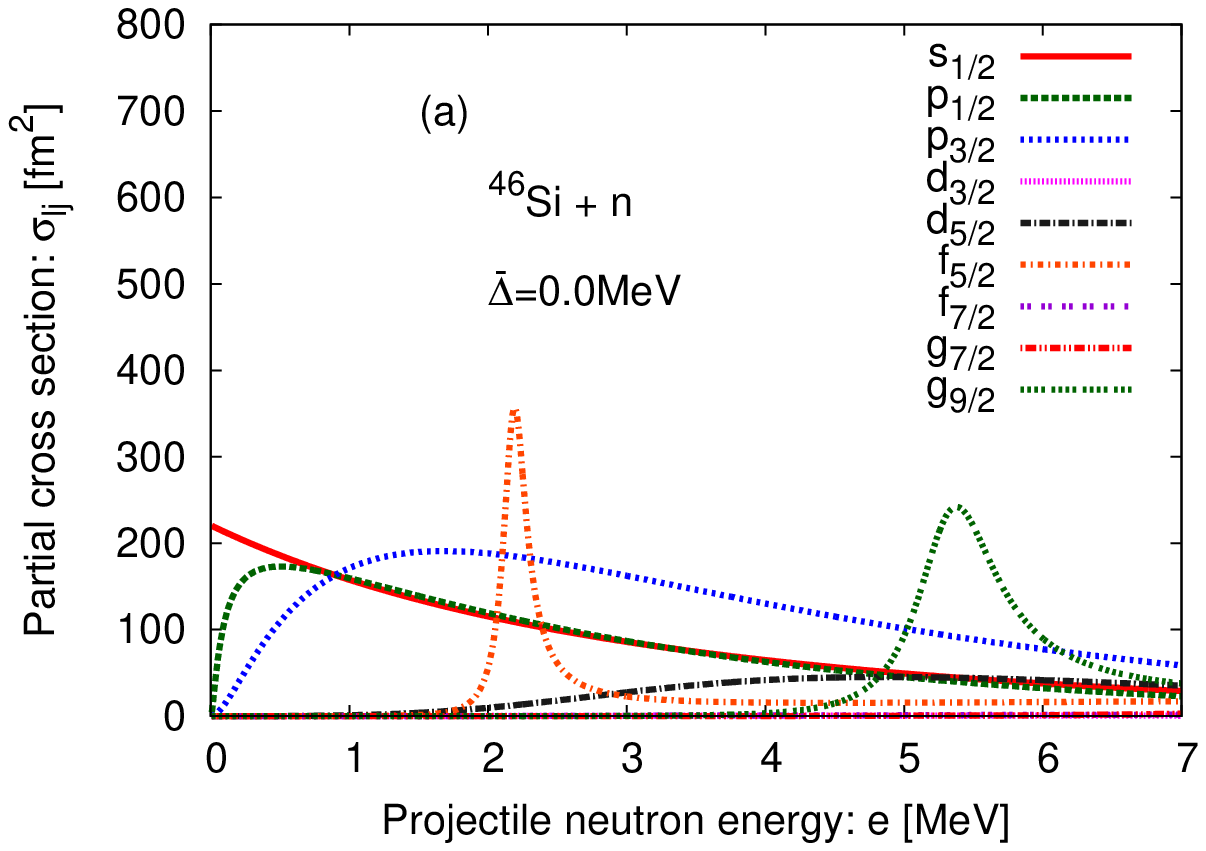}
\includegraphics[scale=0.59]{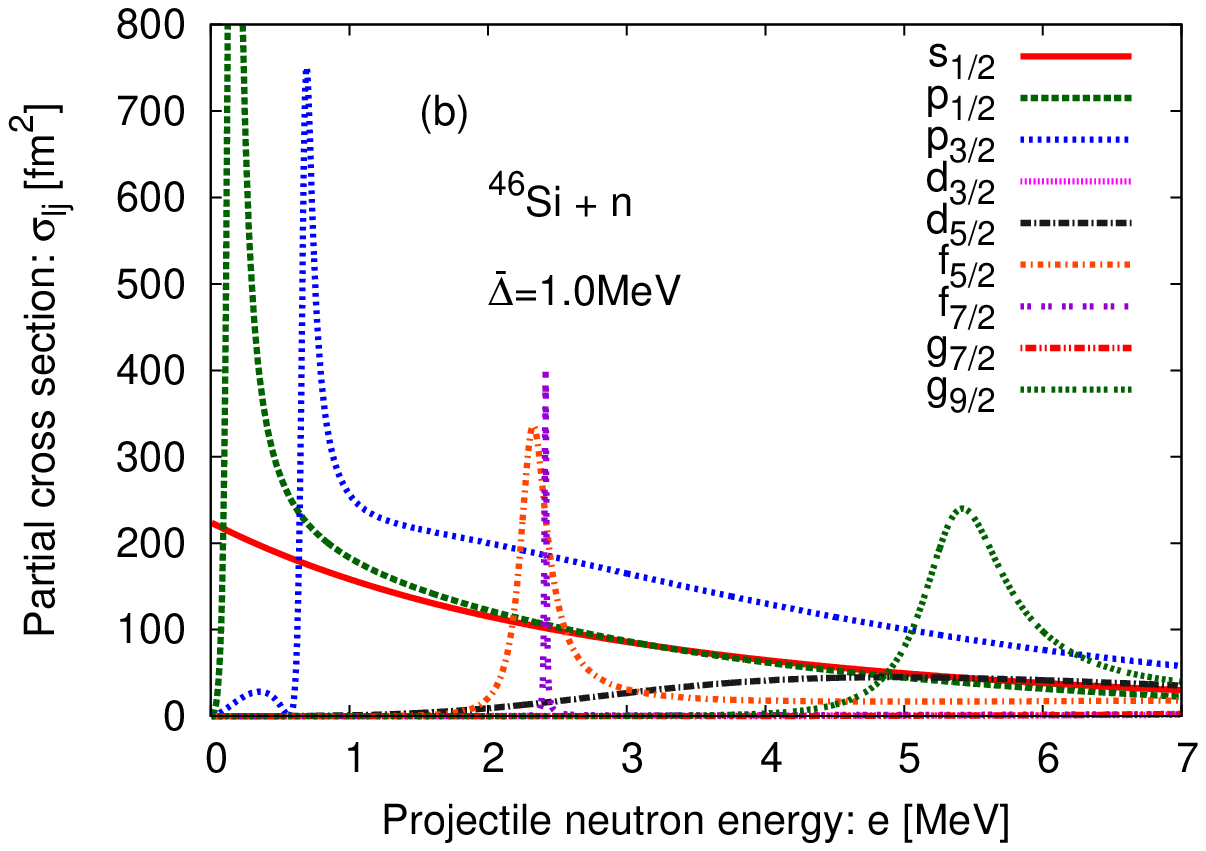}
\caption{(a) Elastic cross sections $\sigma_{lj}$ for various partial waves in the case of $\bar{\Delta}=0.0$ MeV. (b) The same as (a), but in the case of $\bar{\Delta}=1.0$ MeV.} 
\end{center}
\end{figure}

On the other hand, in the case of $\bar{\Delta}$=1.0 MeV, we see narrow low-lying peaks in the $p_{1/2}$, $p_{3/2}$ and $f_{7/2}$ waves, which do not exist in the case of $\bar{\Delta}$=0.0 MeV. These peaks are not potential resonances caused by the centrifugal barrier. These characteristic resonances are the quasi-particle resonances which are caused by the pairing correlation. They are associated with the weakly bound single-particle orbits $2p_{1/2}$, $2p_{3/2}$ and $1f_{7/2}$. With $\bar{\Delta}=1.0$ MeV, the quasi-particle states corresponding to $2p_{3/2}$ or $2p_{1/2}$ orbits become unbound resonances, seen as the low-lying peaks in Fig.~1~(b). It is noted the $2p_{1/2}$ resonance energy is lower than that of $2p_{3/2}$, with the ordering opposite to the standard single-particle states.

In the following discussion, we focus on the low-lying 2$p_{1/2}$ resonance. Figure~2 shows the elastic cross sections and the phase shifts of the 2$p_{1/2}$ resonance which are obtained for various values of the pairing strength $\bar{\Delta}$. It is seen in these figures that the resonance is influenced significantly by the pairing strength $\bar{\Delta}$.
\begin{figure}
\begin{center}
\includegraphics[scale=0.59]{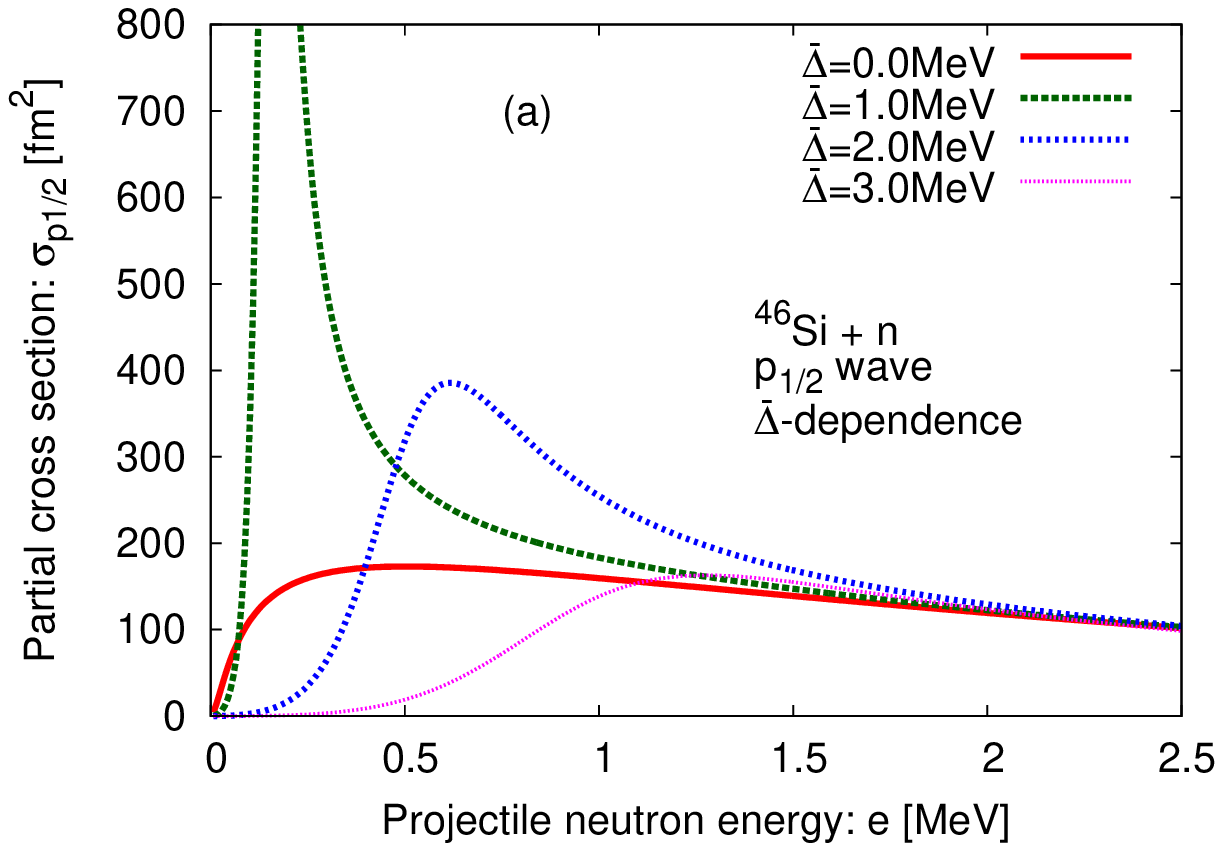}
\includegraphics[scale=0.59]{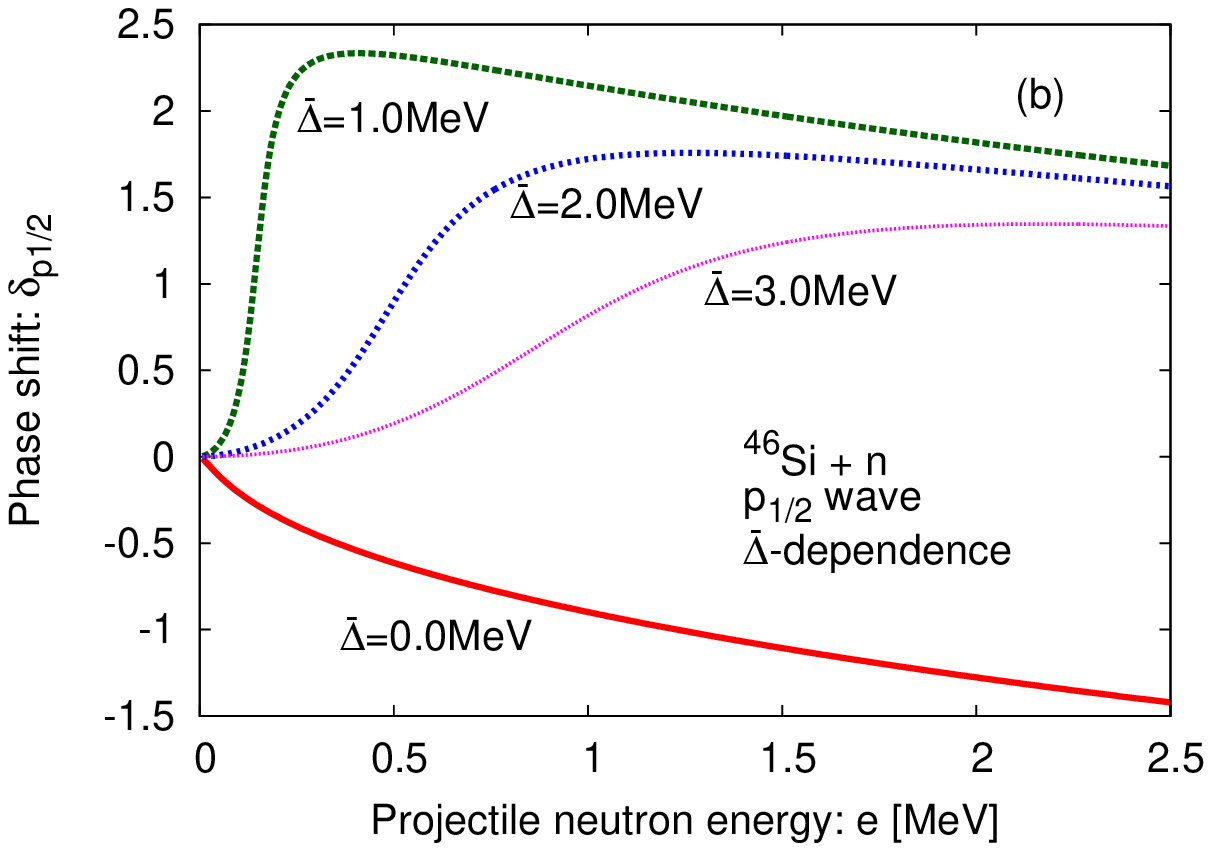}
\caption{(a) Elastic cross section $\sigma_{p1/2}$ of the partial wave $p_{1/2}$ for various values of $\bar{\Delta}$. (b) Elastic phase shift $\delta_{p1/2}$ of the partial wave $p_{1/2}$ for various values of $\bar{\Delta}$.} 
\end{center}
\end{figure}

For $\bar{\Delta}=0.0$ MeV, no single-particle resonance is seen in the $p_{1/2}$ wave since the $2p_{1/2}$ orbit is bound with the single-particle energy $e_{2p1/2}=-0.056$ MeV and the corresponding quasi-particle energy $E_{2p1/2}=|e_{2p1/2}-\lambda|=0.213$ MeV is smaller than the threshold $|\lambda|=0.269$ MeV. As $\bar{\Delta}$ increases ($\bar{\Delta}\sim 0.5$ MeV), the $2p_{1/2}$ quasi-particle state acquires the quasi-particle energy $E$ larger than $|\lambda|$, and then appears in the continuum region as a resonance. For further increasing $\bar{\Delta}\gtrsim1$ MeV, both the resonance width and the resonance energy are found to increase. The increase of the resonance energy may be anticipated qualitatively as the conventional BCS expression for the quasi-particle energy $E=\sqrt{(e_{\mathrm{sp}}-\lambda)^{2}+\Delta^{2}}$ suggests. The increase of the width $\Gamma$ as the function of the pair potential ($\propto |\bar{\Delta}|^{2}$) is suggested in the perturbative analysis ~\cite{Belyaev1987,Bulgac1980}. However, we found that non-trivial pairing effects are involved here as we discuss below.

\subsection{Resonance width and resonance energy}
We evaluate the resonance width and the resonance energy in order to investigate quantitatively effects of the pairing correlation on these values. We extract the resonance width and the resonance energy from the calculated phase shift using a fitting method. We employ the following function to fit:
\begin{equation}
\delta(e)=\arctan \left( \frac{2(e-e_{R})}{\Gamma} \right)+a(e-e_{R})+b
\label{fiteq_sec6}
\end{equation}
where $e$, $\Gamma$ and $e_{R}$ are the kinetic energy of the scattering neutron, the resonance width (defined as the full width at half maximum (FWHM)) and the resonance energy, respectively, and constants $a$ and $b$ representing a smooth background. We perform the fitting in the following two steps. First, we introduce a tentative energy interval and perform a fitting. Next, using a zero-th order values $e^{(0)}_{R}$ and $\Gamma^{(0)}$, we perform the second fitting for the interval $\max(e^{(0)}_{R}-\Gamma^{(0)}, 0)\le e\le e^{(0)}_{R}+\Gamma^{(0)}$. Figure~3 shows the resonance width $\Gamma$ and the resonance energy $e_{R}$ for various values of $\bar{\Delta}$ corresponding to Fig.~2~(b). The vertical axis is the resonance width $\Gamma$ and the horizontal axis is the resonance energy $e_{R}$. Both the resonance width $\Gamma$ and the resonance energy $e_{R}$ increase as the strength of pairing correlation $\bar{\Delta}$ increases. Although the resonance width $\Gamma$ becomes larger than the resonance energy $e_{R}$ for $\bar{\Delta}\geq 2.0$ MeV, we regard it as a meaningful resonance since the fitting has as good quality as that in the cases of $\bar{\Delta}<2.0$ MeV.
\begin{figure}[t]
\begin{center}
\includegraphics[scale=0.59]{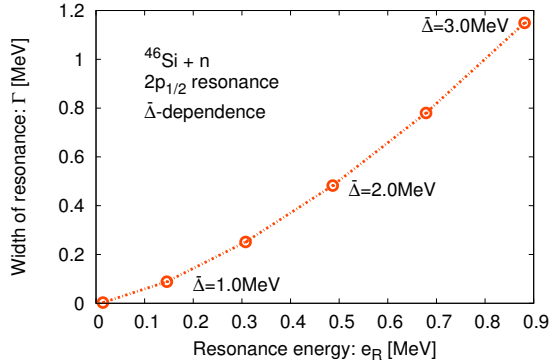}
\caption{The $e_{R}$-$\Gamma$ relation of the $2p_{1/2}$ quasi-particle resonance for various values of $\bar{\Delta}$. The vertical axis is the resonance width $\Gamma$ and the horizontal axis is the resonance energy $e_{R}$.} 
\end{center}
\end{figure}

To investigate systematically influence of the position of single-particle orbit on the resonance, we change not only the strength of pairing correlation $\bar{\Delta}$ but also the single-particle energy of the $2p_{1/2}$ orbit. We vary the depth of the Woods-Saxon potential $V_{0}$ to change the single-particle energy. The variation from the original value is denoted by $\Delta V_{0}$. Figure~4~(a) shows the $2p_{1/2}$ single-particle energy as a function of $\Delta V_{0}$. The length of vertical bars in the figure represents the resonance width (FWHM). It is seen that the $2p_{1/2}$ orbit enters into the continuum as the depth is arisen by $\Delta V_{0}\sim 0.5$ MeV. The resonance width (vertical bars) grows with further raise of potential depth. The height of centrifugal barrier $E_{\mathrm{barrier}}$ for the $p_{1/2}$ wave (the dotted curve in the Fig.~4~(a)) is $\sim$0.5 MeV, being independent approximately on $\Delta V_{0}$. Figure~4~(b) shows the $e_{R}$-$\Gamma$ relation of the single-particle potential resonance corresponding to the Fig.~4~(a). For $\Delta V_{0}\gtrsim 4.0$ MeV, the resonance width is very broad, $\Gamma\gtrsim 2e_{R}$, as expected from $e_{R}\gtrsim E_{\mathrm{barrier}}$.

\begin{figure}
\begin{center}
\includegraphics[scale=0.59]{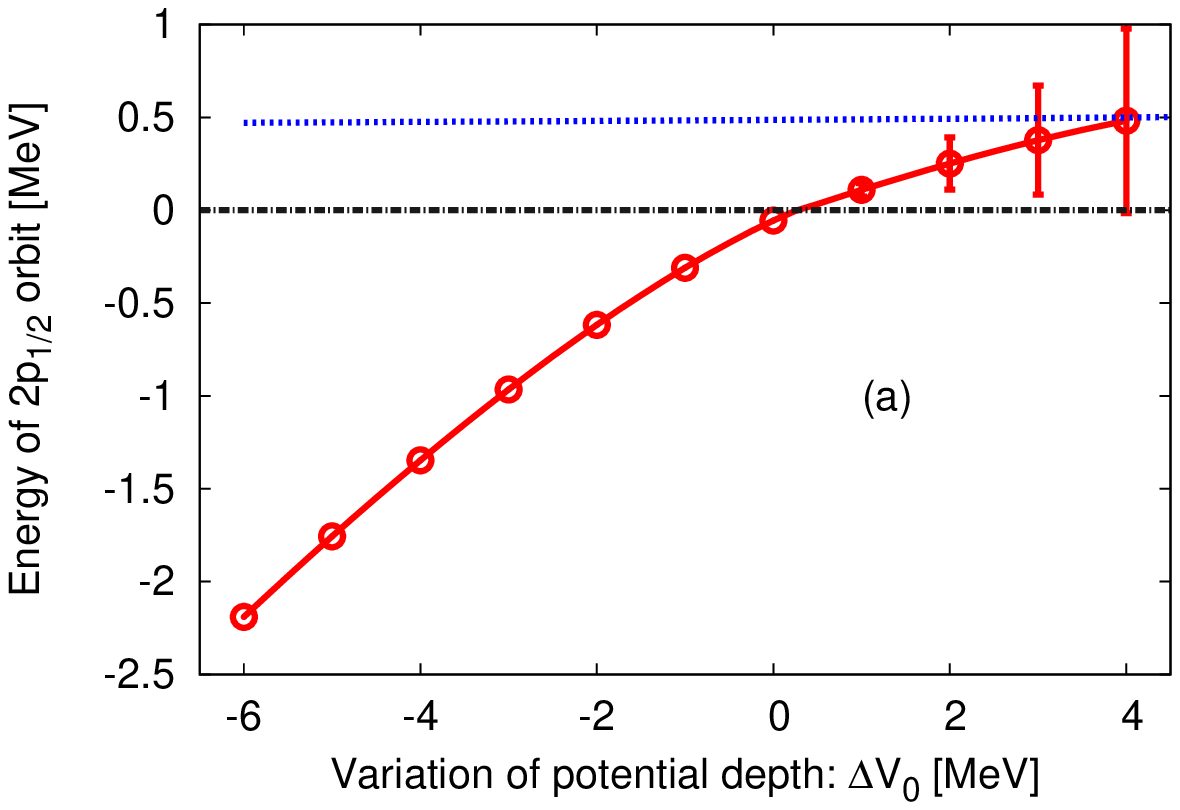}
\includegraphics[scale=0.59]{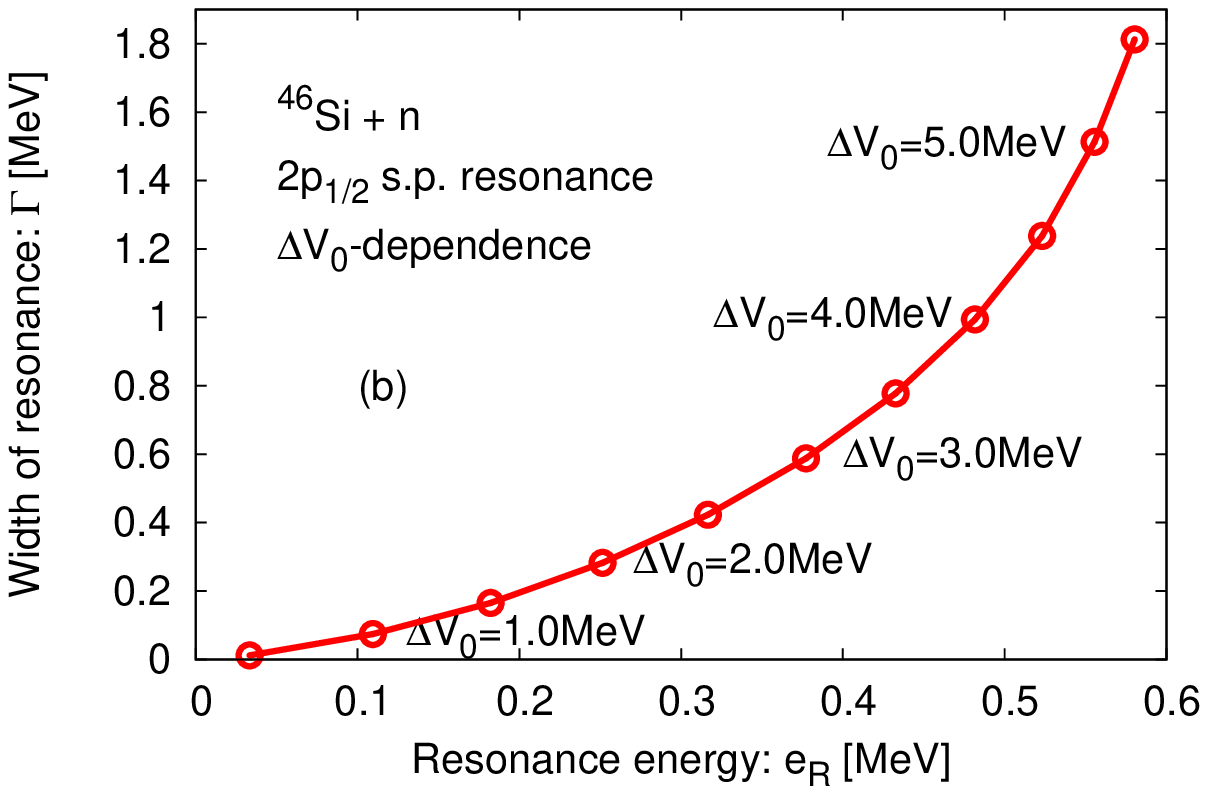}
\caption{(a) The single-particle energy of the neutron $2p_{1/2}$ orbit for various depths of the Woods-Saxon potential. The vertical axis is the neutron single-paticle energy and the horizontal axis is the variation of potential depth $\Delta V_{0}$. Positive single-particle energy represents the resonance energy, and the length of attached vertical bar represents the resonance width (FWHM). The dotted line indicates the height of the centrifugal barrier. (b) The $e_{R}$-$\Gamma$ relation of the $2p_{1/2}$ single-particle potential resonance for various potential depths $\Delta V_{0}$.} 
\end{center}
\end{figure}

The resonance width and the resonance energy evaluated for various $\bar{\Delta}$ and $\Delta V_{0}$ are plotted in the $e_{R}$-$\Gamma$ plane in Fig.~5. As a reference, the $e_{R}$-$\Gamma$ relation of the single-particle potential resonance (Fig.~4~(b)) is also shown.
\begin{figure}
\begin{center}
\includegraphics[scale=0.59]{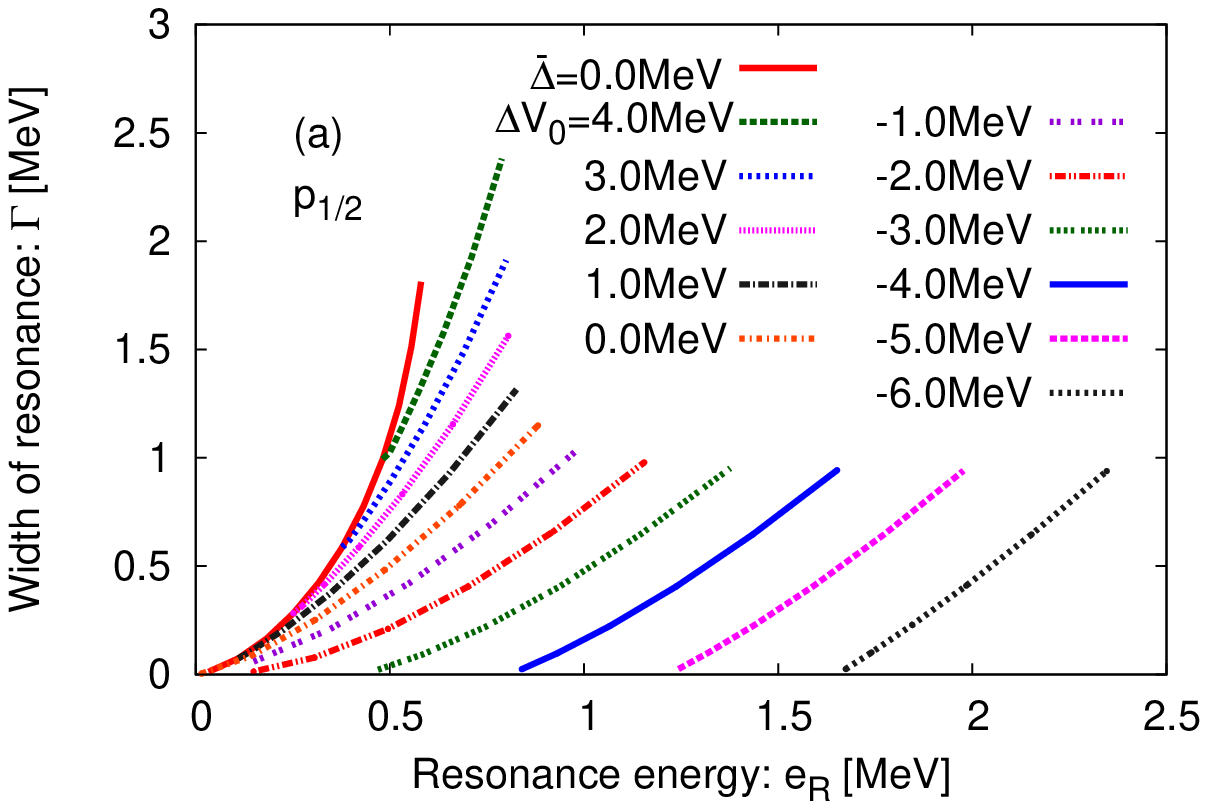}
\includegraphics[scale=0.59]{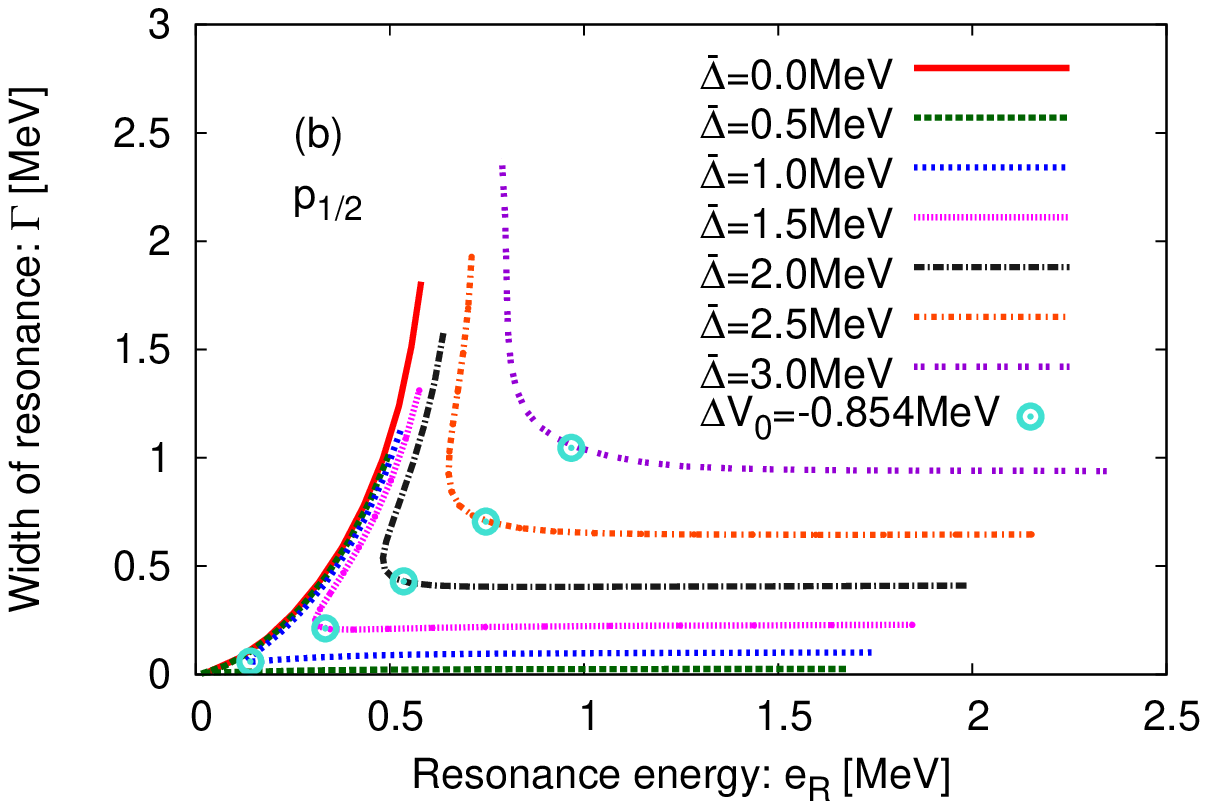}
\caption{The $e_{R}$-$\Gamma$ relation of the $2p_{1/2}$ quasi-particle resonance with various values of $\bar{\Delta}$ and $\Delta V_{0}$. (a) The $e_{R}$-$\Gamma$ relation for given values of $\Delta V_{0}$ with varying $\bar{\Delta}$ from $0.0$  to $3.0$ MeV. (b) The $e_{R}$-$\Gamma$ relation for given values of $\bar{\Delta}$ with varying $\Delta V_{0}$ from $-6.0$ to $4.0$ MeV. The curve with $\bar{\Delta}=0.0$ MeV is the $e_{R}$-$\Gamma$ relation of the $2p_{1/2}$ single-particle resonance, shown in Fig. 4 (b).} 
\end{center}
\end{figure}

Figure~5~(a) is a plot displaying dependence of $\Gamma$ on $\bar{\Delta}$ for fixed values of $\Delta V_{0}$. We see that both the resonance width and the resonance energy increase with increasing $\bar{\Delta}$ for all the values of $\Delta V_{0}$. Figure~5~(b) is another plot showing dependence on $\Delta V_{0}$ for fixed values of $\bar{\Delta}$. A distinctive feature seen in Fig.~5 is that the quasi-particle resonance exist even at energies $e_{R}$ higher than the barrier height $E_{\mathrm{barrier}}\sim0.5$ MeV. It is seen also that the $e_{R}$-$\Gamma$ relation displays two different features. One is seen in the bottom-right region of Fig.~5~(b) where the resonance width changes only slightly for change of the resonance energy. The other is that the resonance width increases sensitively as the resonance energy changes, seen in the upper-left region. This difference in the $e_{R}$-$\Gamma$ relation is related to whether the $2p_{1/2}$ orbit is located above or below the Fermi energy. In other words, the difference originates from whether the original $2p_{1/2}$ orbit is particle-like or hole-like. More precisely, the $2p_{1/2}$ orbit is particle-like (hole-like) for $\Delta V_{0}>-0.854$ MeV ($\Delta V_{0}\leq -0.854$ MeV). The boundary $\Delta V_{0}=-0.854$ MeV is plotted in Fig.~5~(b) with open circles. In the following discussion, we call the former a particle-like quasi-particle resonance, and the latter a hole-like quasi-particle resonance.

Concerning the hole-like quasi-particle resonance, the resonance width approximately independent on the resonance energy $e_{R}$. Deviation from this simple behavior is seen for $e_{R}\lesssim1.0$ MeV. As for the particle-like quasi-particle resonance, the behavior is much more complicated and non-trivial. We shall examine these points in the following subsections.

\subsection{Pairing effect on the hole-like quasi-particle resonance}
Let us first analyze the hole-like quasi-particle resonances, i.e. in the case of $e_{{\rm sp}} < \lambda$. As already seen in connection with Fig.~5~(b), the dependence of the resonance width $\Gamma$ on the average pairing gap $\bar{\Delta}$ appears rather simple:  $\Gamma$ increases monotonically with $\bar{\Delta}$ while $\Gamma$ depends only weakly on the resonance energy $e_R$ or the single-particle energy $e_{{\rm sp}}$. We shall now analyze the pairing dependence of the resonance width $\Gamma$ by comparing with the analytical expression~\cite{Belyaev1987,Bulgac1980} which is derived for the hole-like quasi-particle resonance on the basis of the perturbation with respect to the pairing gap  or the pairing potential. 

The perturbative evaluation assumes that a single-hole state with energy $e_{{\rm sp}}$ and wave function $\varphi_i(\vec{r}\sigma)$ couples to unbound single-particle states $\varphi_{e}(\vec{r}\sigma)$ only weakly via the pair potential $\Delta(\vec{r})$. This leads to the expression
\begin{equation}
\Gamma_{i}=2\pi\left| \sum_{\sigma}\int d\vec{r}\varphi^{\dagger}_{i}(\vec{r\sigma})\Delta(\vec{r})\varphi_{e}(\vec{r}\sigma) \right|^2 \propto \left|\Delta_{\mathrm{average}}\right|^{2}
\label{qpreswidth}
\end{equation}
where the wave function of the unbound single-particle orbit at energy $e$ is normalized as
\begin{equation}
\sum_{\sigma}\int d\vec{r}\varphi^{\dagger}_{e}(\vec{r}\sigma)\varphi_{e^{\prime}}(\vec{r}\sigma)=\delta (e - e^{\prime}).
\label{kikaku}
\end{equation}
The resonance energy in the zero-th order is $e_R^0= |e_i - \lambda| + \lambda=|e_i|- 2|\lambda|$, corresponding to
the quasi-particle energy $E_i^0 = |e_i - \lambda| $ of the hole state.

We shall now compare the resonance width $\Gamma$ obtained from the numerical fit to the phase shift and that from the perturbative evaluation Eq.~(\ref{qpreswidth}). The results are shown in Fig.~6, which plots the evaluated widths as functions of the average pairing gap  $\bar{\Delta}$. The perturbative calculation using Eq.~(\ref{qpreswidth}) is performed in two different ways, and they are plotted with the upward and downward triangles in Fig.~6. The curve with upward triangles is the case where the wave functions $\varphi_{i}$ and $\varphi_{e}$ of the hole and continuum orbits are fixed, and only $\Delta (r)$ is changed. For the energy of $\varphi_{e}$, we use the zero-th order resonance energy $e^{0}_{R}=|e_{2p1/2}|-2|\lambda|$. This scheme is named ``Fermi's golden rule 1'' hereafter. In the calculation for the curve with downward triangles, we fix the single-particle wave function of bound orbit $\varphi_{i}$, but we choose the energy $e$ of $\varphi_{e}$ that reproduces the resonance energy $e_{R}(\bar{\Delta})$ obtained from the phase shift for each $\bar{\Delta}$ (called ``Fermi's golden rule 2'').

Figure~6~(a) shows the $\bar{\Delta}$-dependence of resonance width $\Gamma$ for the resonance arising from the $2p_{1/2}$ hole state at $e_{\mathrm{sp}}=-4.127$ MeV ($\Delta V_{0}=-10.0$ MeV). Figure~6~(b) and (c) are the same as (a), but these are for the $2p_{1/2}$ hole orbits at $e_{\mathrm{sp}}=-1.347$ MeV  ($\Delta V_{0}=-4.0$ MeV) and $e_{\mathrm{sp}}=-0.618$ MeV  ($\Delta V_{0}=-2.0$ MeV), respectively. Figure~6~(a) is the case where the single-particle energy of hole orbit is smaller than the Fermi energy $\lambda =-0.269$ MeV by about 4 MeV. This is a typical hole-like quasi-particle resonance since the resonance width $\Gamma$ evaluated with perturbative calculations reproduce the non-perturbative evaluation of the resonance width $\Gamma$. Deviations from the perturbative expression are seen in Fig.~(b) and (c). The difference between the perturbative and the non-perturbative evaluation becomes large as the single-particle energy $e_{\mathrm{sp}}$ approaches the Fermi energy $\lambda$ and the pair potential grows as seen in Fig.~6~(b) and (c).
\begin{figure}[t]
 \centering
 \begin{minipage}{0.3\hsize}
  \begin{center}
   \includegraphics[width=50mm,angle=0]{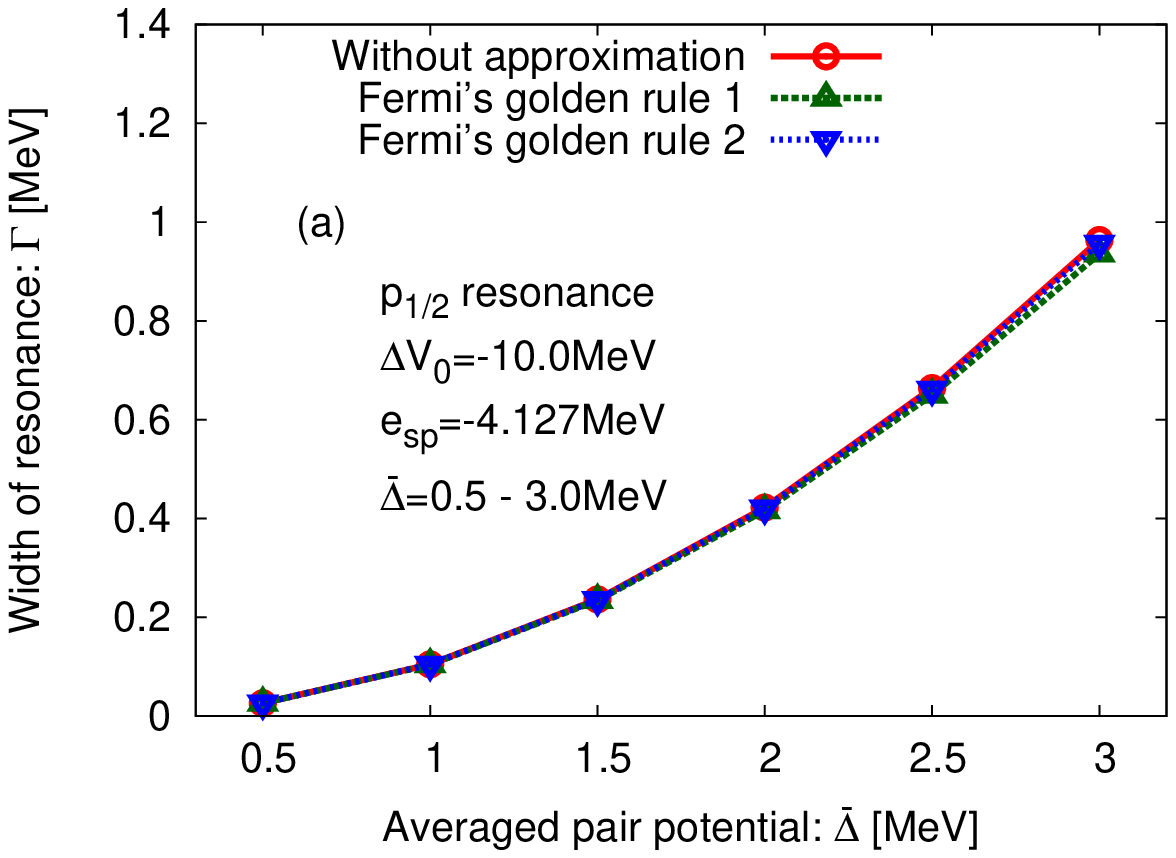}
  \end{center}
 \end{minipage}
 \begin{minipage}{0.3\hsize}
  \begin{center}
   \includegraphics[width=50mm,angle=0]{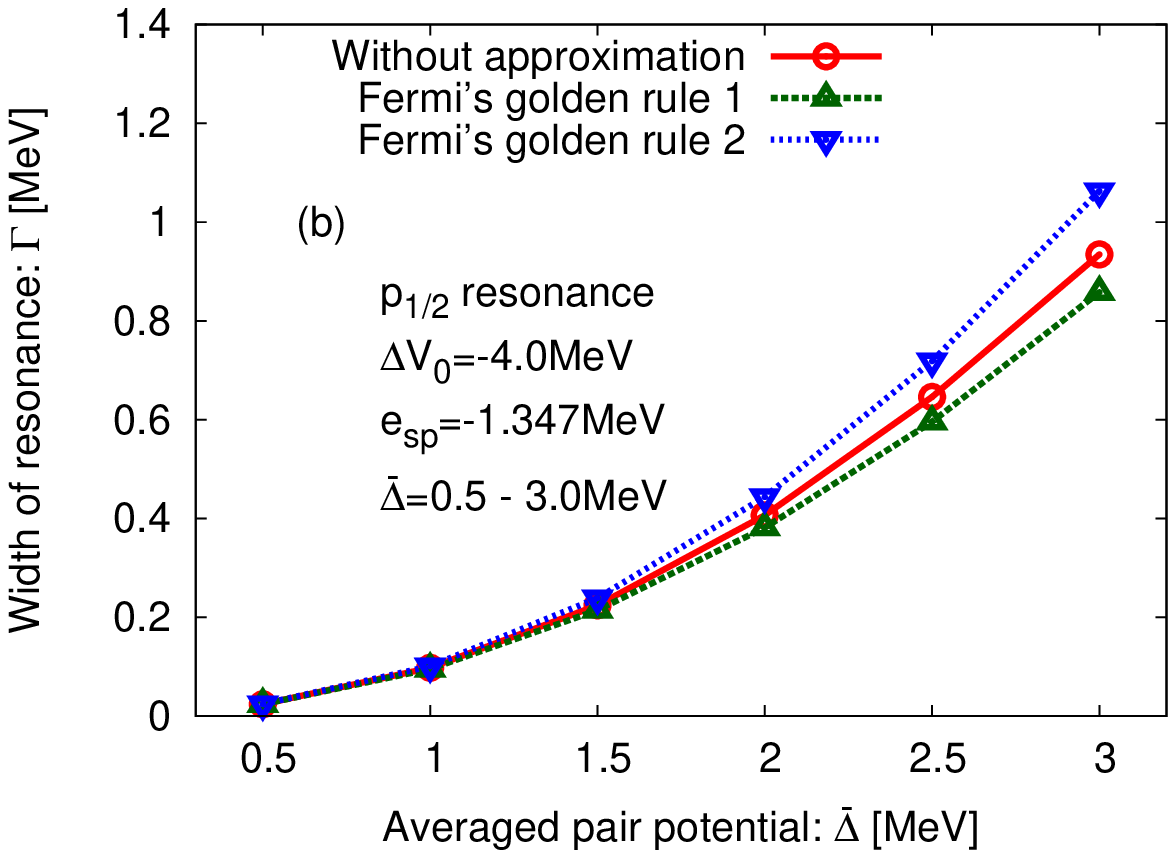}
  \end{center}
 \end{minipage}
 \begin{minipage}{0.3\hsize}
  \begin{center}
   \includegraphics[width=50mm,angle=0]{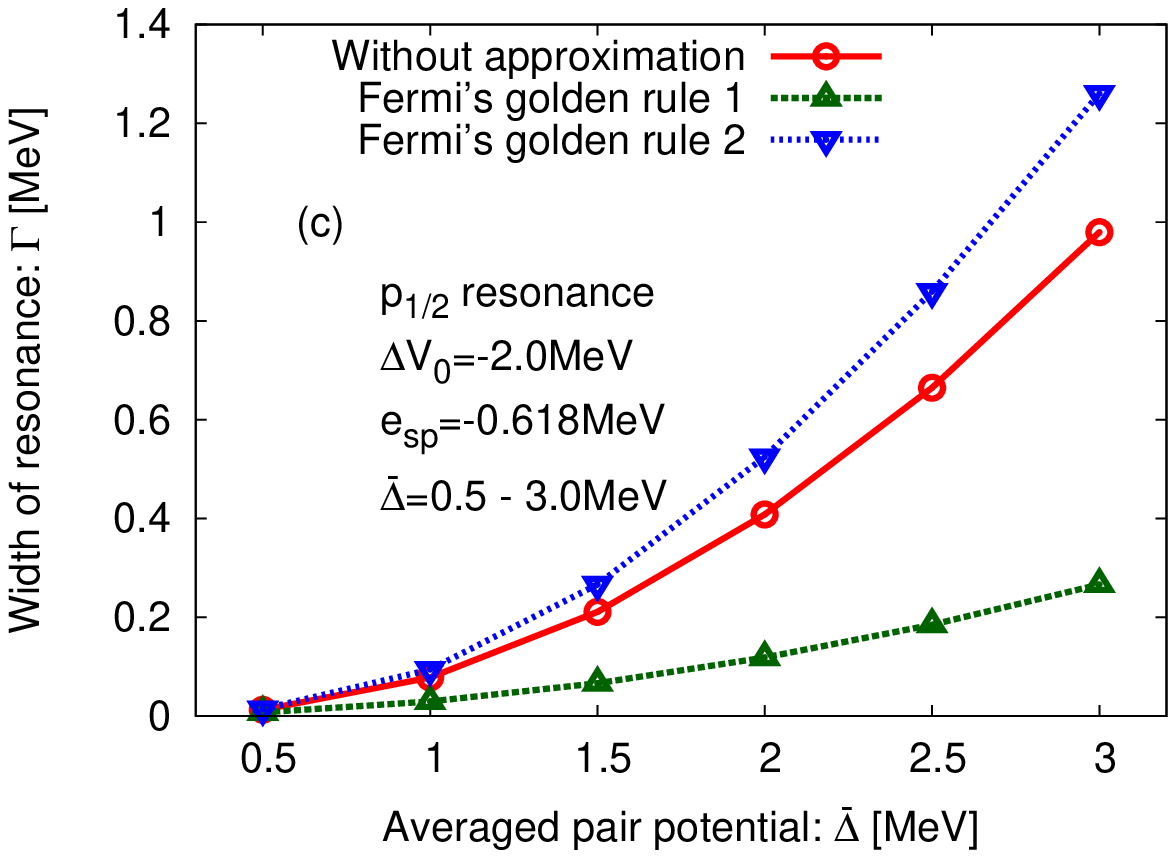}
  \end{center}
 \end{minipage}
\caption{Comparison of the perturbative evaluations of the resonance width $\Gamma$ obtained with Eq.~(10) (plotted with triangles) and the width $\Gamma$ obtained from the phase shift (plotted with circles), for $2p_{1/2}$ hole-like quasi-particle resonance, corresponding to the single-particle energies $e_{\mathrm{sp}}=-4.127$ MeV ($\Delta V_{0}=-10.0$ MeV) [panel (a)], $-1.347$ MeV ($\Delta V_{0}=-4.0$ MeV) [panel (b)] and $-0.618$ MeV ($\Delta V_{0}=-2.0$ MeV) [panel (c)]. The horizontal axis is the average pairing potential $\bar{\Delta}$. The upward triangle is the perturbative width $\Gamma$ in the scheme ``Fermi's golden rule 1'', while the downward triangle is that in the scheme ``Fermi's golden rule 2'' (see text).}
\end{figure}

\begin{figure}[t]
 \centering
 \begin{minipage}{0.3\hsize}
  \begin{center}
   \includegraphics[width=50mm,angle=0]{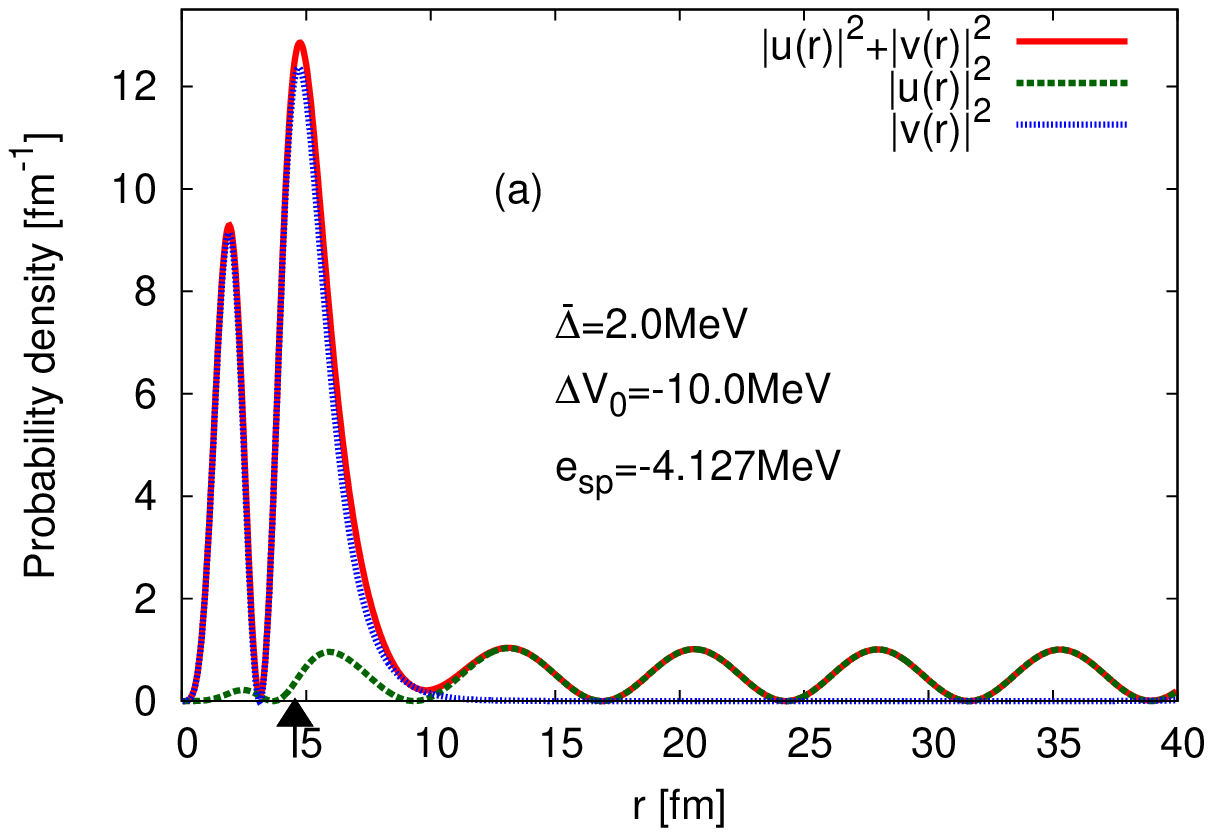}
  \end{center}
 \end{minipage}
 \begin{minipage}{0.3\hsize}
  \begin{center}
   \includegraphics[width=50mm,angle=0]{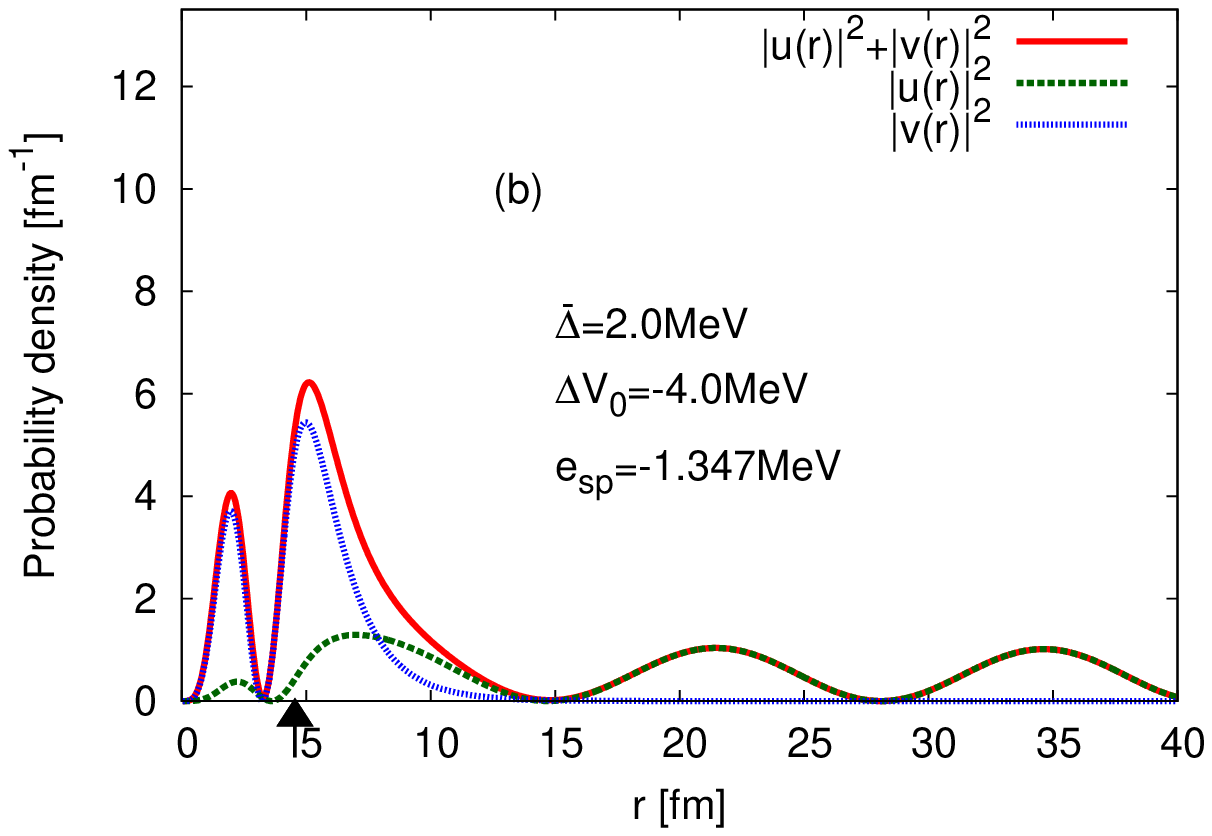}
  \end{center}
 \end{minipage}
 \begin{minipage}{0.3\hsize}
  \begin{center}
   \includegraphics[width=50mm,angle=0]{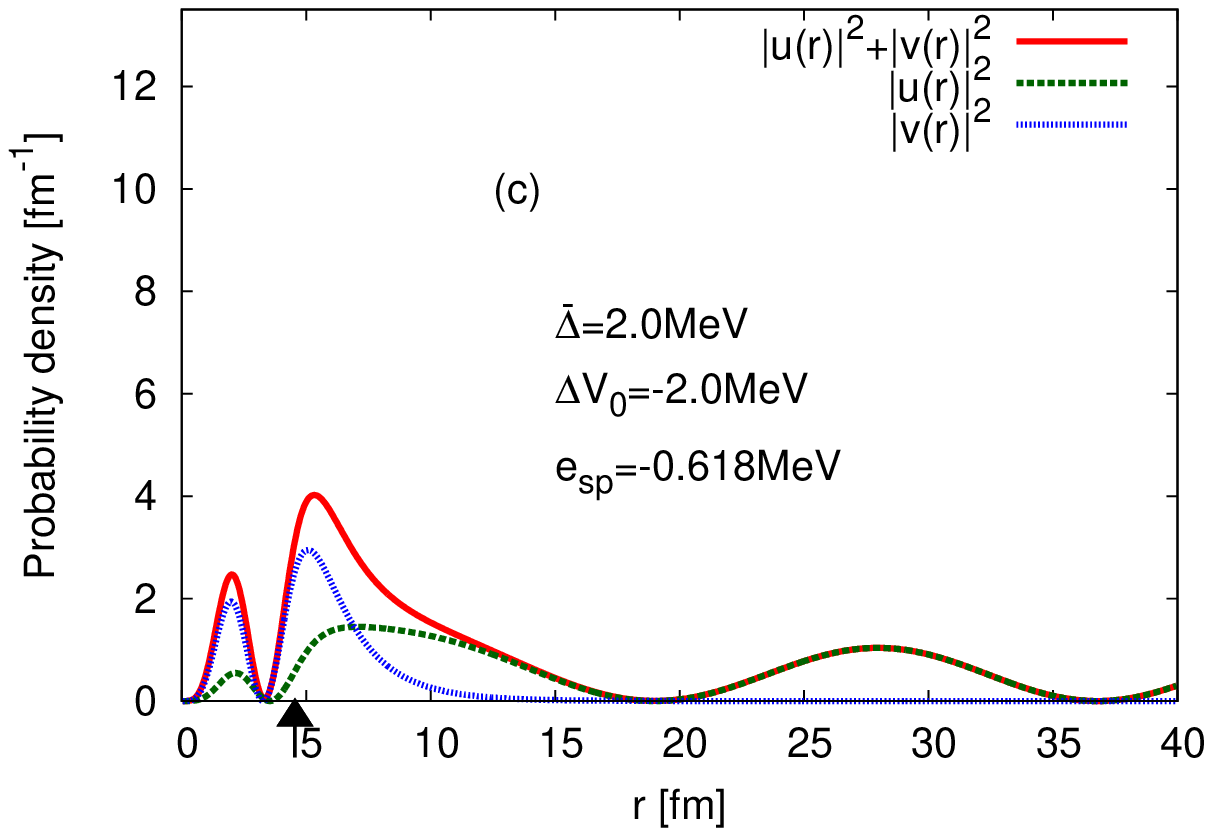}
  \end{center}
 \end{minipage}
\caption{Probability distribution $|u(r)|^{2}+|v(r)|^{2}$ of the $2p_{1/2}$ quasi-particle resonance, corresponding to (a) $e_{\mathrm{sp}}=-4.127$ MeV ($\Delta V_{0}=-10.0$ MeV), (b) $e_{\mathrm{sp}}=-1.347$ MeV ($\Delta V_{0}=-4.0$ MeV) and (c) $e_{\mathrm{sp}}=-0.618$ MeV ($\Delta V_{0}=-2.0$ MeV). The pairing strength is commonly $\bar{\Delta}=2.0$ MeV. Partial probabilities $|u(r)|^{2}$ and $|v(r)|^{2}$ associated with the particle- and hole-components, respectively, are also plotted. The Woods-Saxon radius $R=4.550$ fm is indicated with an arrow. The wave functions $u(r)$ and $v(r)$ are normalized so that $u(r)$ has a common asymptotic amplitude 1.}
\end{figure}

Figure~7 shows the probability distributions $|v(r)|^2$ and $|u(r)|^2$ of the three examples of the hole-like quasiparticle resonance. The panels (a), (b) and (c) correspond to Fig. 6~(a), (b) and (c), respectively (for $\bar{\Delta}=2.0$ MeV). Note that $|u(r)|^{2}$ is the probability distribution of the particle-component while $|v(r)|^{2}$ is that of the hole-component, and $|u(r)|^{2}+|v(r)|^{2}$ is the total probability to find the quasi-particle at position $r$. As expected, the probability $|u(r)|^{2}$ of the particle-component is much smaller than the probability $|v(r)|^{2}$ of the main hole-component in  the case (a) where the perturbation works well. Contrarily, in the case (c) where the perturbation breaks down, $|u(r)|^{2}$ is comparable to the probability $|v(r)|^{2}$ of the main hole-component indicating strong mixing of the particle-component. For more quantitative argument we evaluate the probability distributions $|v(r)|^2$ and $|u(r)|^2$ integrated within the nuclear surface: $\bar{u}^{2}=\int^{R}_{0}|u(r)|^{2}dr$ and $\bar{v}^{2}=\int^{R}_{0}|v(r)|^{2}dr$, and evaluate the ratio $\bar{u}^{2}/\bar{v}^{2}$. The ratio is $0.021$ and $0.254$ for the case (a) and (c), respectively. In the case (b), corresponding to the boundary region for the breaking down of the perturbation, the ratio is $0.091$. 

We have examined the applicability of the perturbative evaluation, Eq.~(\ref{qpreswidth}), systematically for all the combinations of $\bar{\Delta}$ and $\Delta V_{0}$ shown in Fig.~5. We adopt a criterion that both of the two evaluations of Eq.~(\ref{qpreswidth}) with different choices of $\varphi_{e}$ agree with the non-perturbative numerical evaluation of the resonance width within 10\% error. We find then that the applicability of Eq.~(\ref{qpreswidth}) is represented in terms of the single-particle energy $e_{\mathrm{sp}}$, the Fermi energy $\lambda$ and the pair gap $\bar{\Delta}$ as
\begin{equation}
e_{\mathrm{sp}}\lesssim \lambda -0.5\bar{\Delta}.
\end{equation}

We also examined validity of Eq.~(10) in terms of the ratio $\bar{u}^{2}/\bar{v}^{2}$. It is found that the applicability of Eq.~(10) is represented also by
\begin{equation}
\bar{u}^{2}/\bar{v}^{2}\lesssim 0.1.
\label{uv}
\end{equation}

The above analysis indicates that the perturbative evaluation works not only for the quasi-particle resonances associated with deeply-bound hole orbit, which has been considered previously~\cite{Belyaev1987,Bulgac1980}, but also for quasi-particle resonances arising from a shallowly-bound hole orbit, for instance, that with $e_{\mathrm{sp}}\sim\lambda-0.5\bar{\Delta}$. Even in the latter case, the mixing of the particle-component into the main hole-component is small $\bar{u}^{2}\lesssim0.1\bar{v}^{2}$. This is probably the reason why the perturbation works in the rather broad situation. On the contrary, it is natural that the perturbation, Eq.~(\ref{uv}), breaks down in the case of $e_{\mathrm{sp}}>\lambda$, where the dominant component of the quasi-particle state is not the hole-component $v(r)$, but the particle-component $u(r)$. A quite different, probably non-perturbative, mechanism of the pairing effect on the resonance width is expected in this case.

\subsection{Pairing effect on the particle-like quasi-particle resonance}
We then analyze the particle-like quasi-particle resonances, i.e. those in the case of $e_{{\rm sp}} \geq \lambda$. As typical examples, we examine two cases with $e_{2p1/2}=-0.056$ MeV ($\Delta V_{0}=0.0$ MeV) and with $e_{2p1/2}=0.251$ MeV ($\Delta V_{0}=2.0$ MeV). Note $e_{2p1/2}>\lambda$ in both cases. Curves in Fig.~5~(a) corresponding to these cases are shown in Fig.~8. The $e_R$-$\Gamma$ relation of the single-particle potential resonance is also shown as a reference.
\begin{figure}[t]
\begin{center}
\includegraphics[width=75mm,angle=0]{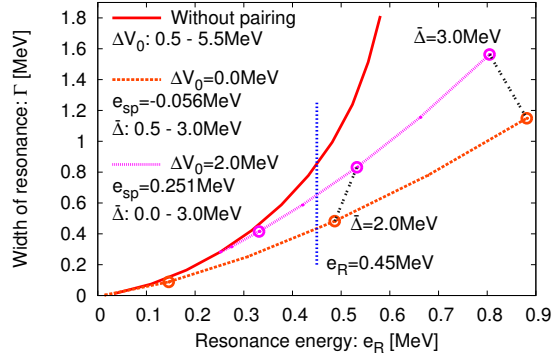}
\end{center}
\caption{The $e_{R}$-$\Gamma$ relation of the $2p_{1/2}$ quasi-particle resonance in the case of particle-like single-particle energy $e_{{\rm sp}}=-0.056$ MeV ($\Delta V_{0}=0.0$ MeV) and $e_{{\rm sp}}=0.251$ MeV ($\Delta V_{0}=2.0$ MeV) (dashed and dotted curves), obtained for varying the average pairing gap $\bar{\Delta}=0.0 - 3.0$ MeV. The $e_{R}$-$\Gamma$ relation of the $2p_{1/2}$ single-particle potential resonance is also shown (solid curve).}
\label{sienewid_p12}
\end{figure}

As seen in Fig.~8 (and also in Fig.~5~(a)), increase of the pairing potential increases monotonically both the resonance width $\Gamma$ and the resonance energy $e_{R}$, displaying a trend similar to that of the hole-like quasi-particle resonance. However, Fig.~5~(b) indicates also that increase of the resonance energy with a fixed value of the pair potential leads to the increase of the resonance width  in the particle-like case. We therefore suppose that two mechanisms are involved here. One is a kinematical effect: Due to the increase of the resonance energy, the penetrability of the centrifugal barrier increases, and consequently it leads to the increase of $\Gamma$. The other is a direct pairing effect, originating from the mixing among the particle- and hole-component caused by the pair potential.

In order to extract the latter mixing effect, we compare these three curves at the same resonance energy. As an example, we make a comparison at $e_{R}=0.45$ MeV. We then find that the resonance width for $\bar{\Delta}=1.634$ MeV is narrower than that for $\bar{\Delta}=0.0$ MeV and the width for $\bar{\Delta}=1.897$ MeV is the smallest among the three cases. The resonance widths for these three cases are listed in Table~3, together with other examples compared at $e_{R}=0.300$ and $0.375$ MeV. It shows that the pairing correlation has an effect to {\it reduce} the resonance width if the comparison is made at the same resonance energy.
\begin{table}[t]
  \centering
  \begin{tabular}{cccccccccccc} \hline
    $e_{R}$ [MeV] & \multicolumn{3}{c}{0.300} && \multicolumn{3}{c}{0.375} && \multicolumn{3}{c}{0.450}\\ \hline
    $\bar{\Delta}$ [MeV] & 0.0 & 0.728 & 1.477 && 0.0 & 1.246 & 1.688 && 0.0 & 1.634 & 1.897 \\
    $\Gamma$ [MeV] & 0.387 & 0.361 & 0.244 && 0.582 & 0.500 & 0.338 && 0.854 & 0.652 & 0.453 \\
    $e_{\mathrm{sp}}$ [MeV] & 0.300 & 0.251 & -0.056 && 0.375 & 0.251 & -0.056 && 0.450 & 0.250 & -0.056 \\ \hline
  \end{tabular}
\caption{Resonance width $\Gamma$ of the $2p_{1/2}$ quasi-particle and single-particle resonances which have $e_{R}=0.300, 0.375$ and 0.450 MeV for three different values of $\bar{\Delta}$. The single-particle resonance energy (or bound single-particle energy) $e_{\mathrm{sp}}$ is also listed.}
\label{enewid_comp}
\end{table}

To examine mechanism of the reduced resonance width, we look into wave functions of the three resonances with $e_{R}=0.450$ MeV. Figure~9 shows the probability distribution of the resonant quasi-particle states with $e_{R}=0.450$ MeV. In the case of $\bar{\Delta}=0.0$ MeV, the hole-component $v(r)$ vanishes and $u(r)$ coincide with the single-particle wave function of the $2p_{1/2}$ potential resonance. With finite values of $\bar{\Delta}$, and increasing of $\bar{\Delta}$, the probability $|u(r)|^{2}+|v(r)|^{2}$ within the surface of the nucleus ($r\lesssim R$) become larger. This is consistent with our finding that the resonance width become narrower with larger pair potential. In particular, it is seen that the increase of the probability inside the nucleus originates mainly from the increase of the hole-component $v(r)$.
\begin{figure}[t]
 \centering
 \begin{minipage}{0.3\hsize}
  \begin{center}
   \includegraphics[width=50mm,angle=0]{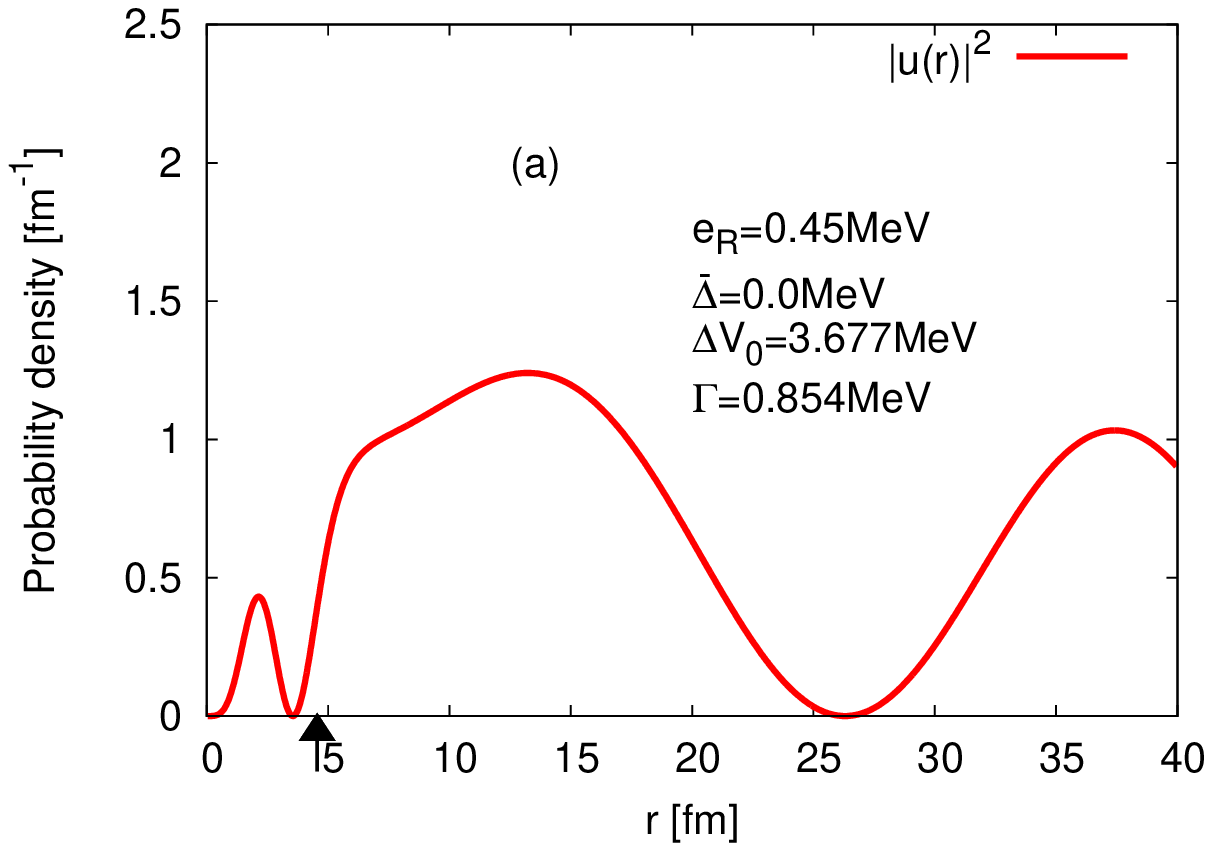}
  \end{center}
 \end{minipage}
 \begin{minipage}{0.3\hsize}
  \begin{center}
   \includegraphics[width=50mm,angle=0]{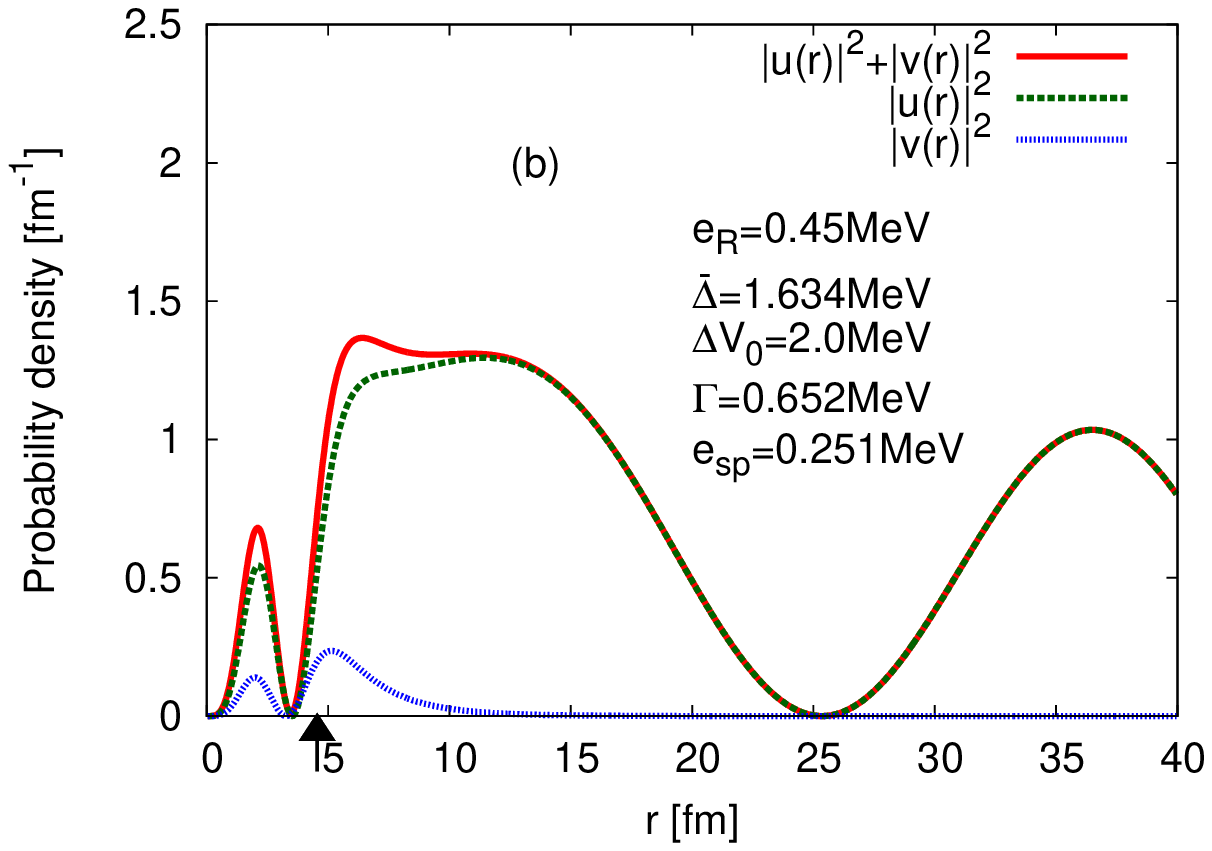}
  \end{center}
 \end{minipage}
 \begin{minipage}{0.3\hsize}
  \begin{center}
   \includegraphics[width=50mm,angle=0]{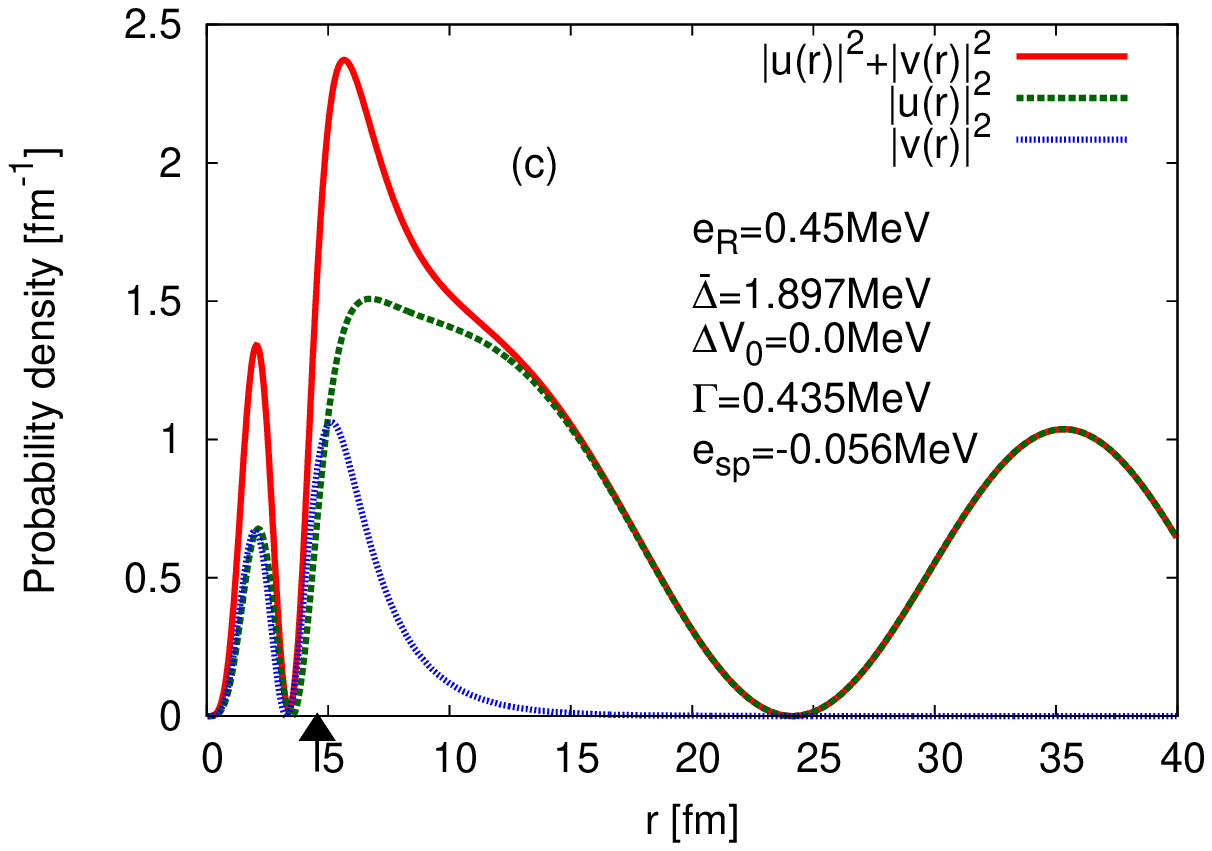}
  \end{center}
 \end{minipage}
\caption{Probability distribution of the $2p_{1/2}$ resonances with common resonance energy $e_{R}=0.45$ MeV, but for different pairing strengths: (a) $\bar{\Delta}=0.0$ MeV, (b) $\bar{\Delta}=1.634$ MeV and (c) $\bar{\Delta}=1.897$ MeV.}
\end{figure}

\begin{table}[t]
  \centering
  \begin{tabular}{cccccccccccc} \hline
    $e_{R}$ [MeV] & \multicolumn{3}{c}{0.300} && \multicolumn{3}{c}{0.375} && \multicolumn{3}{c}{0.450}\\ \hline
    $\bar{\Delta}$ [MeV] & 0.0 & 0.728 & 1.477 && 0.0 & 1.246 & 1.688 && 0.0 & 1.634 & 1.897 \\
    $\bar{v}^{2}/\bar{u}^{2}$ & 0.0 & 0.069 & 0.891 && 0.0 & 0.187 & 1.003 && 0.0 & 0.297 & 1.107 \\
    $v^{2}_{\mathrm{BCS}}/u^{2}_{\mathrm{BCS}}$ & 0.0 & 0.045 & 0.456 && 0.0 & 0.107 & 0.503 && 0.0 & 0.161 & 0.543 \\ \hline
  \end{tabular}
\caption{The ratio $\bar{v}^{2}/\bar{u}^{2}$ of the probability distributions of the hole- and particle-components of the quasi-particle wave functions of the $2p_{1/2}$ resonance, evaluated for different values of $\bar{\Delta}$, but for the common resonance energy $e_{R}$. The $v^{2}_{\mathrm{BCS}}/u^{2}_{\mathrm{BCS}}$ based on the BCS formula is also listed. See text for details.}
\label{prob_part_comp}
\end{table}

The increase of the hole-component $v(r)$ is a natural consequence of the pairing correlation. Here we recall the simple BCS formula for the $u$ and $v$ factors: the amplitudes of the particle- and hole-components are
\begin{equation}
v^{2}_{\mathrm{BCS}}=\frac{1}{2}\left( 1-\frac{e-\lambda}{E} \right),\quad u^{2}_{\mathrm{BCS}}=\frac{1}{2}\left( 1+\frac{e-\lambda}{E} \right),
\label{bcs}
\end{equation}
respectively, with the quasi-particle energy $E=\sqrt{(e-\lambda)^{2}+\Delta^{2}}$. The hole-probability $v^{2}_{\mathrm{BCS}}$, which vanishes for $\Delta =0$, increases with increasing $\Delta$ since the pair potential causes the mixing among the particle- and hole-components. We consider that a similar mixing mechanism takes place in the present case. We show in Table~3, the ratio $\bar{v}^{2}/\bar{u}^{2}$ of the particle- and hole-components obtained from the HFB calculation, and $v^{2}_{\mathrm{BCS}}/u^{2}_{\mathrm{BCS}}$ evaluated by using the BCS formula (\ref{bcs}). Here the quasi-particle energy $E$ is related to the resonance energy $e_{R}$ as $E=|\lambda|+e_{R}$. It is seen that the increasing trend of $\bar{v}^{2}/\bar{u}^{2}$ is consistent with that of the BCS formula except a difference by a factor of $\sim0.5$. The consistency is also seen in examples at the other resonance energies.

The above observation leads to the following interpretation. The amplitude $v(r)$ of hole-component increases due to the mixing of the hole- and particle-components via the pair potential. Since the hole-component $v(r)$ is localized inside and around the nuclear surface, the increase of $v(r)$ leads to the increase of probability distribution $|u(r)|^{2}+|v(r)|^{2}$ inside the nuclear radius $r\lesssim R$. This brings about the decrease of the resonance width.

As a secondary mechanism, we find that the particle-component $u(r)$ inside and around the surface increases with $\bar{\Delta}$. This also contributes to the increase of  $|u(r)|^{2}+|v(r)|^{2}$. We will leave analysis of this mechanism for forthcoming paper since this contribution is small compared with the contribution from the hole-component.

\section{Conclusion}
The quasi-particle resonance is predicted in the Bogoliubov's quasi-particle theory as an unbound single-particle mode of excitation caused by the pair correlation in nuclei. Expecting strong influence of the pair correlation, we have studied in the present paper properties of the quasi-particle resonance emerging in nuclei near the neutron drip-line. We focused on the resonance in the $p$ wave neutron with low kinetic energy in the ${}^{46}$Si + n system, and analyzed in detail how the pair correlation controls the width of the quasi-particle resonance.

By solving numerically the Hartree-Fock-Bogoliubov equation in the coordinate space to obtain the quasi-particle wave function satisfying the scattering boundary condition, we calculate the phase shift of the neutron elastic scattering and then extract the resonance energy and the resonance width. Analyses are performed systematically for various strengths of the average pairing gap, and for different situations concerning whether the quasi-particle state is particle-like or hole-like, i.e. whether the single-particle orbit being the origin of the resonance is located above or below the Fermi energy.

We have disclosed that the pairing effect on the width of the particle-like quasi-particle resonance is very different from that of the hole-like quasi-particle resonance, for which a perturbative treatment~\cite{Belyaev1987,Bulgac1980} of the pair potential is known. A peculiar feature of the particle-like quasi-particle resonance is that the resonance width for a strong pairing is smaller than that of a weaker pairing if comparison is made at the same resonance energy: The pairing correlation has an effect to {\it reduce} the resonance width. This is opposite to the pairing effect on the of the hole-like quasi-particle resonance. In the hole-like case, the pair potential causes a coupling of the hole state to the scattering neutron states, leading to a decay of the hole state. In the particle-like case, in contrast, the pair potential causes the scattering state, represented by the particle-component $u(r)$ of the quasi-particle wave function, to mix with the hole-component $v(r)$, which is however confined inside and around the nuclear surface. Therefore, with increasing the strength of the pair potential, the probability of the quasi-particle state inside the nucleus increases, and hence the width (decay probability) decreases.

Concerning the hole-like quasi-particle resonances, we have examined the applicability of the perturbative evaluation~\cite{Belyaev1987,Bulgac1980} of the resonance width. It is found that the perturbation can be applied not only to the quasi-particle resonances associated with deeply bound hole state, as known previously, but also to hole-like quasi-particle resonances whose corresponding hole energy is close to the Fermi energy $\lambda$. More precisely the applicability condition is evaluated to be $e_{{\rm sp}} \lesssim \lambda - 0.5\bar{\Delta}$.

\begin{acknowledgments}
We thank Kenichi Yoshida for useful discussions. This work is supported by Grant-in-Aid for Research Fellowships of Japan Society for the Promotion of Science (JSPS) for Young Scientists. It is also supported by Grant-in-Aid for Scientific Research from Japan Society for Promotion of Science No. 23540294, No. 24105008 and No. 26400268.
\end{acknowledgments}

\end{document}